\documentclass{aa}
\usepackage{txfonts}
\usepackage{graphicx}
\usepackage{natbib}

\begin{document}

\title{{\it Chandra} Observations of Nuclear X-ray Emission from a Sample of Radio Sources}

\author{J. K. Gambill \inst{1}
\and R. M. Sambruna \inst{1,}\inst{2}
\and G. Chartas \inst{3}
\and C. C. Cheung \inst{4}
\and L. Maraschi \inst{5}
\and F. Tavecchio \inst{5}
\and C. M. Urry \inst{6}
\and J. E. Pesce \inst{2}}

\institute{School of Computational Sciences, MS 5C3, George Mason
 University, 4400 University Drive, Fairfax, VA 22030 \and Department
 of Physics and Astronomy, MS 3F3, George Mason University, 4400
 University Drive, Fairfax, VA 22030 \and Department of Astronomy and
 Astrophysics, 525 Davey Lab, The Pennsylvania Stat
 University, State
 College, PA 16802 \and Department of Physics, MS
 057, Brandeis University, Waltham, MA 02454 \and Osservatorio Astronomico di Brera, via
 Brera 28, Milan I-20121, Italy \and Department of
 Astronomy, P.O. Box 208101, Yale University, New Haven, CT 06520}

\offprints{J. K. Gambill, \email{jessica@physics.gmu.edu}}

\date{Received: 14 November 2002 / Accepted: 30 January 2003}

\abstract{We present the X-ray properties of a sample of 17 radio
sources observed with the {\it Chandra} X-ray Observatory as part of a
project aimed at studying the X-ray emission from their radio jets. In
this paper, we concentrate on the X-ray properties of the unresolved
cores. The sample includes 16 quasars (11 core-dominated and 5
lobe-dominated) in the redshift range $z$=0.30--1.96, and one
low-power radio-galaxy at $z$=0.064.  No diffuse X-ray emission is
present around the cores of the quasars, except for the nearby
low-power galaxy that has diffuse emission on a scale and with a
luminosity consistent with other FRIs. No high-amplitude, short-term
variability is detected within the relatively short {\it Chandra}
exposures.  However, 1510$-$089 shows low-amplitude flux changes with
a timescale of $\sim$25 minutes.  The X-ray spectra of the quasar
cores are generally well described by a single power law model with
Galactic absorption. However, in six quasars we find soft X-ray excess
emission below 1.6 keV. Interestingly, we detect an Fe K-shell
emission line, consistent with fluorescent K$\alpha$ emission from
cold Iron, in one lobe- and two core-dominated sources. The average
X-ray photon index for the quasars in the sample is
$\Gamma_{sample}=1.66$ and dispersion $\sigma_{sample}=0.23$. The
average spectral slope for our sample is flatter than the slope found
for radio-quiet quasars and for radio-loud AGNs with larger jet
orientations; this indicates that beaming affects the X-ray emission
from the cores in our sample of quasars.  }

\maketitle
\section{Introduction}

An important distinction in the quasar class is between radio-quiet
and radio-loud active galactic nuclei (AGNs), based on their
optical-to-radio flux ratios (e.g., Kellermann et al. 1989). While the
ultimate source of power is thought to be the same for both classes of
AGNs (i.e., accretion of gas onto a supermassive black hole; Antonucci
1993; Urry \& Padovani 1995), there are subtle but systematic
differences in the continuum and line properties from radio to
optical/UV wavelengths between the classes (Sanders et al. 1989; Miley
\& Miller 1979; Yee \& Oke 1978).  X-ray observations of both types of
sources probe the inner regions and can thus help discriminate the
origin of the radio-loud/quiet difference. However, while radio-quiet
quasars have been extensively studied at X-rays (e.g., George et
al. 2000; Fiore et al. 1998; Nandra et al. 1997a,b), the X-ray
properties of radio-loud AGNs are relatively less well known (e.g.,
Sambruna, Eracleous, \& Mushotzky 2002a and references therein). The
study of the nuclear X-ray emission from radio-loud quasars with
previous X-ray satellites was generally hampered by poor angular
resolution and/or sensitivity.

Earlier studies of AGNs at medium-soft X-rays with {\it Einstein} and
{\it ROSAT} showed that radio-loud quasars have significantly flatter
continuum slopes, $\Gamma\sim$ 1.5, than their radio-quiet
counterparts, $\Gamma\sim$ 2.0 (Fiore et al. 1998; Nandra et
al. 1997a,b; Shastri et 1991; Wilkes \& Elvis 1987).  This was
attributed to dilution of the core emission by a beamed X-ray
component from the jet. Recent systematic studies of radio-loud AGNs
at harder X-rays with {\it ASCA} confirmed that radio-loud quasars are
flatter, $\Gamma\sim$ 1.6, than radio-quiet quasars (Reeves \& Turner
2000). However, observations of Broad-Line Radio Galaxies (BLRGs) and
double-lobe quasars, believed to be at larger jet orientations,
indicate steeper slopes ($\Gamma\sim 1.8$) more similar to Seyferts
(Hasenkopf et al. 2002; Sambruna, Eracleous, \& Mushoztky 1999).

In low-luminosity radio-quiet AGNs, Fe K-shell emission is a
relatively well known property. In several Seyfert 1s a prominent
K$\alpha$ fluorescent emission line from cold Fe was detected with
{\it ASCA} at rest-frame energies 6.4 keV, with Equivalent Widths
(EWs) $\sim$ 250 eV (Nandra et al. 1997b). However, the presence and
the nature of Fe K emission at higher luminosities is not yet
established. Nandra et al. (1997a) detected an Fe line in a
radio-quiet quasar at energies higher than 6.4 keV, consistent with
emission from highly ionized Iron. A highly ionized Fe line, observed
by {\it ASCA} with energies consistent with emission from Fe XXV-XXVI,
may be common in radio-quiet quasars (Reeves \& Turner 2000). An
ionized Fe line was recently confirmed in a high-redshift quasar with
{\it XMM-Newton} (Reeves et al. 2001).

However, Fe K emission is rare in radio-loud AGNs. So far, a weak and
generally narrow Fe K$\alpha$ emission line was observed in the
brightest BLRGs (see Sambruna, Eracleous, \& Mushotzky 2002a and
references therein), and more recently in 3 high-luminosity
(L$_{2-10~keV} \sim 10^{45}$ ergs~s$^{-1}$) radio-loud quasars
(Hasenkopf et al. 2002). While there is evidence that Fe K emission
and other reflection features in BLRGs are intrinsically weak (e.g.,
Eracleous, Sambruna, \& Mushotzky 2000), an additional difficulty in
their detection is the presence of diffuse X-ray emission from the
host galaxy/associated cluster of galaxies, which may dilute any
intrinsic (weak) Fe emission lines from the active nucleus in
low-resolution observations.

The advent of the {\it Chandra} X-ray Observatory, with unprecedented
angular resolution (0.492$^{\prime\prime}$/pixel) and improved
sensitivity, makes significant progress possible in the study of the
X-ray emission from the cores of radio-loud quasars. With {\it
Chandra}, the X-ray emission from the nucleus can be isolated from the
extended (kiloparsec-scale) jet emission and other extended components
(e.g., thermal radiation from a cluster). Also, the X-ray spectral
index and other features in the spectrum can be better quantified.

In this paper, we report on the X-ray properties of the nuclei of 17
radio-loud sources observed with {\it Chandra}. The sample was
selected based on the properties of the extended radio features as
part of a survey aimed at finding the jet X-ray and optical
counterparts with {\it Chandra} and the {\it Hubble Space Telescope} ({\it
HST}). The {\it Chandra} and multiwavelength jet properties of the
first six observed sources were discussed in Sambruna et al. (2002b),
while the remaning sources will be presented in a future
publication. Here we concentrate on the X-ray properties of the cores
in our {\it Chandra} images.

The plan of the paper is as follows: Section~2 presents the sample, the
{\it Chandra} observations, and the analysis methods, Section~3 describes
the results of the spatial, timing, and spectral analysis, and Section~4
discusses the implications of Section~3. Throughout this work, H$_0=75$ km
s$^{-1}$ Mpc$^{-1}$ and $q_0=0.5$ are adopted.

\section{Observations}
\begin{table*}[]
\begin{center}
\begin{tabular}{llccccc}
\multicolumn{7}{l}{{\bf Table 1: Sample Sources and their Basic Properties}} \\
\multicolumn{7}{l}{   } \\ \hline \hline
& & & & & & \\
\multicolumn{1}{c}{Source} & \multicolumn{1}{c}{Alt Name} & m$_{V}$ & $z$ & N$_{H}^{Gal}$ & Type & R$_i$  \\ 
\multicolumn{1}{c}{(1)} & \multicolumn{1}{c}{(2)} & \multicolumn{1}{c}{(3)} & (4) & (5) & (6) & (7) \\
& & & & & & \\ \hline
& & & & & & \\
0405$-$123 & PKS & 14.8 & 0.574 &  3.81 & FSRQ & 1.1  \\
& & & & & & \\
0605$-$085 & PKS & 18.5 & 0.870 & 21.0 & FSRQ & 14.5  \\
& & & & & & \\
0723+679 & 3C 179 & 18.0 & 0.846 & 4.31 & SSRQ & 0.44 \\
& 4C +67.14 & & & & & \\
0802+103 & 3C 191 & 18.4 & 1.956 & 2.28 & SSRQ & 0.03 \\
& 4C +10.25 &  & & & & \\
0836+299 & 4C +29.30 & 15.7 & 0.064 & 4.06 & RG & 0.06 \\  
& & & & & & \\
0838+133 & 3C 207 & 18.5 & 0.684 & 4.00 & SSRQ & 0.43 \\
& 4C +13.38 & & & & & \\ 
1040+123 & 3C 245 & 17.3 & 1.029 & 2.87 & SSRQ & 1.1$^a$ \\
& 4C +12.37 & & & & & \\
1055+018 & 4C +01.28 & 18.3 & 0.888 & 3.40 & FSRQ/BL & 19.2  \\
& & & & & & \\
1136$-$135 & PKS & 16.1 & 0.554 & 3.59 & SSRQ &  0.29\\
& & & & & & \\
1150+497 & 4C +49.22 & 17.1 & 0.334 & 2.05 & FSRQ & 2.4\\ 
& & & & & & \\
1354+195 & 4C +19.44 & 16.0 & 0.720 & 2.18 & FSRQ & 2.7 \\ 
& & & & & & \\
1510$-$089 & PKS & 16.5 & 0.361 & 7.96 & FSRQ & 6.7 \\
& &  & & & & \\
1641+399 & 3C 345 & 16.0 & 0.594 & 1.13 & FSRQ & 9.1  \\
& 4C +39.48 &  & & & & \\
1642+690 & 4C +69.21 & 19.2 & 0.751 & 4.54 & FSRQ & 6.8  \\
& & & & & &\\
1741+279 & 4C +27.38 & 17.7 & 0.372 & 4.59 & FSRQ & 1.6 \\
& & & & & & \\
1928+738 & 4C +73.18 & 16.5 & 0.302 & 7.71 & FSRQ & 16.8 \\
& & & & & & \\
2251+134 & 4C +13.85 & 19.5 & 0.673 & 4.98 & FSRQ & 1.1 \\
& & & & & & \\ \hline
\end{tabular}
\end{center}
{\bf Columns Explanation:} {1=Source IAU name; 2=Common
source names; 3=Optical magnitude, V-band; 4=Redshift; 5=Galactic
column density in units $10^{20}$ cm$^{-2}$ from NED Galactic
Extinction Calculator; 6=Source type. FSRQ: Flat Spectrum Radio
Quasar; SSRQ: Steep Spectrum Radio Quasar; RG: Radio Galaxy; BL: BL
Lac Object; 7=Ratio of core to extended radio power at 5~GHz,
corrected for the redshift (observed value $\times (1+z)^{-1}$) from
Cheung et al. (2003).  Core-dominated sources have R$_i >$1, while
lobe-dominated sources have R$_i <$1.}

\noindent{\bf Notes:} {$a$=The quasar 1040+123 is known to
be a lobe-dominated source with a variable nucleus (Hough \& Readhead
1989).  However, this core was brighter than it had been previously
during the epoch of our archival observation, hence the $R_i$ value
greater than unity.  We classify this object as a lobe-dominated SSRQ
in our discussion, as previous observations have established (Hough \&
Readhead 1989).}
\end{table*}

\subsection{The Sample}

The selection criteria of the sample of 17 radio-loud sources are
discussed in Sambruna et al. (2002b). Briefly, the sample was
extracted from the list of radio jets of Bridle \& Perley (1984) and
Liu \& Xie (1992) based on the length and brightness of the jet, in
order to match the {\it Chandra} and {\it HST} observing
capabilities. As a result, the sample contains 17 targets and is
biased toward the presence of a bright radio jet. The targets include
16 quasars, 11 Flat Spectrum Radio Quasars (FSRQs, 5 Steep Spectrum
Radio Quasars (SSRQs), and 1 nearby, low-power radio galaxy
(0836+299). All the quasars have one-sided jets indicating that
beaming is likely to be important in these sources.

{\it Chandra} and {\it HST} observations were awarded to us for 16 of
the 17 sources in the sample. {\it Chandra} observations of the 17th
source, 3C 207, were awarded to other investigators (Brunetti et
al. 2002); we complete our sample with archival {\it Chandra} data of
3C~207.  Table 1 lists the sources and their basic properties.  Based
on the listed ratio of core-to-jet radio power (Cheung et al. 2003),
11 sources are classified as core-dominated and 6 as lobe-dominated
(Table~1, Col. 7).

All sources except 0836+299 are high-power radio sources
(P$_{1.4~GHz}$\lower.5ex\hbox{$\; \buildrel > \over \sim\;$}$10^{32}$
ergs~s$^{-1}$) and exhibit a Fanaroff-Riley II (FRII) radio morphology
(Fanaroff \& Riley 1974). The source 0836+299 has low-power
(P$_{1.4~GHz}\sim 3\times 10^{31}$ ergs~s$^{-1}$) and a radio
morphology resembling a Fanaroff-Riley I (FRI) (van Breugel et
al. 1986). We classify this source as an FRI.

\subsection{Data Acquisition, Reduction, and Analysis}

The observations were made with the Advanced CCD Imaging Spectrometer
(ACIS-S), with the sources at the nominal aimpoint on chip S3. The
awarded exposures were 10 ks per target, with occasionally longer or
shorter exposures to accommodate gaps in the {\it Chandra} observing
schedule. To avoid pileup of the core, a 1/8 subarray mode was used
with one operational CCD, reducing the frame time to 0.4 s. Despite
this precaution, the nuclear X-ray emission was so strong that in 10/17
sources pileup was still present. The measured counts/frame for the
piledup sources range from 0.1 to 0.5 counts/frame. Based on Figure
6.25 of the {\it Chandra} Proposer Observer Guide (POG), the pileup fraction
in the sources of our sample ranges between 1\% and 10\%.  Pileup is
expected to produce significant changes to the spectral slopes
($\delta\Gamma < -0.1$) for observed count rates greater than 0.1
counts/frame. 

The {\it Chandra} data were reduced using \verb+CIAO+~v.~2.1.2 and
following standard criteria, using the latest calibration files
provided by the {\it Chandra} X-ray Center (CXC).  Pixel randomization
was removed and only events for {\it ASCA} grades 0, 2--4, and 6 were
retained for the analysis.  The data were further restricted to the
energy range 0.5--8 keV, where the background is negligible and the
ACIS-S calibration is best known.  We also checked that no flaring
background events occurred during the observations. After screening,
the effective exposure times range between 8.4--10.4 ks, with one
longer exposure, 37.5 ks (0838+133). Table 2 lists the details of the
{\it Chandra} observations, where a flag has been added to denote the
sources affected by pileup (Col. 5).

\begin{table*}[]
\begin{center}
\begin{tabular}{lcccclc}
\multicolumn{7}{l}{{\bf Table 2: Log of Chandra Observations }} \\
\multicolumn{7}{l}{   } \\ \hline \hline
& & & & & \\
\multicolumn{1}{c}{Source} & Start Date & Exposure & Net Counts & Pileup &\multicolumn{1}{c}{Box} & Ratio \\
\multicolumn{1}{c}{(1)} & (2) & (3) & \multicolumn{1}{c}{(4)} & (5) & \multicolumn{1}{c}{(6)} & (7) \\ 
& & & & & \\ \hline
& & & & & \\
0405$-$123 & 2001-07-22 & 8661.6 & 10445 & Y & 2.5 $\times$ 20 & 120 \\
0605$-$085 & 2001-05-01 & 8663.2 & 1254 & N & 4.0 $\times$ 3.0 & 16 \\
0723+679 & 2001-01-15 & 9333.3 & 1820 & N & 6.0 $\times$ 3.0 & 54 \\
0802+103 & 2001-03-07 & 8318.9 & 253$^a$ & N & 3.0 $\times$ 1.0 & 0.9 \\
0836+299 & 2001-04-08 & 7690.9 & 80$^a$ & N & 4.0 $\times$ 18 & 0.3 \\
0838+133 & 2000-11-04 & 37,544 & 5872 & Y & 6.0 $\times$ 3.0 & 20 \\
1040+123 & 2001-02-12 & 10402. & 1956 & N & 3.0 $\times$ 4.0 & 92 \\
1055+018 & 2001-01-09 & 9314.3 & 5887 & Y & 2.0 $\times$ 4.0 & 370 \\
1136$-$135 & 2000-11-30 & 8906.2 & 3162 & Y & 5.0 $\times$ 9.0 & 22 \\
1150+497 & 2000-12-10 & 9293.7 & 7612 & Y & 4.0 $\times$ 8.0 & 38 \\
1354+195 & 2001-01-08 & 9055.7 & 5624 & Y & 3.0 $\times$ 30 & 60 \\
1510$-$089 & 2001-03-23 & 9241.4 & 5988 & Y & 6.0 $\times$ 8.0 & 31 \\
1641+399 & 2001-04-27 & 9055.7 & 6076 & Y & 5.0 $\times$ 4.0 & 82\\
1642+690 & 2001-03-08 & 8326.9 & 1382 & N & 4.0 $\times$ 4.5 & 44 \\
1741+279 & 2001-07-21 & 8895.3 & 2224 & Y & 5.0 $\times$ 4.0 & 115 \\
1928+738 & 2001-04-27 & 8392.5 & 6899 & Y & 6.0 $\times$ 4.0 & 87 \\
2251+134 & 2000-10-21 & 9185.6 & 1699 & N & 7.0 $\times$ 7.0 & 183 \\
& & & & & \\ \hline 
\end{tabular}
\end{center}
{\bf Columns Explanation:} {1=Source IAU name; 2=Observation start
date (Year-Month-Day); 3=Net Chandra exposure, in seconds; 4=Core
photon counts in the energy range 0.5--8 keV from a 2$^{\prime\prime}$
apeture; 5=Flag indicating pileup of the Chandra data; 6=Dimensions of
box used to extract the jet count rate, width $\times$ length,
($^{\prime\prime}$); 7=Ratio of core-to-jet counts.}

\noindent{\bf Notes:} {$a$=Core region is defined by a 1.5\arcsec\
extraction radius (see text Sect.~2.2).}
\end{table*}

For the seven sources not affected by pileup (unpiled sources), X-ray
fluxes and spectra were extracted using a circular region with a
2$^{\prime\prime}$ radius centered on the brightest pixel.  Based on
the ACIS-S encircled energy fraction (Fig. 6.3 in the {\it Chandra}
POG), a source extraction radius of 1$^{\prime\prime}$\ encircles
$\; \buildrel > \over \sim$ 90\% of the flux of a point source at 1
keV.  From this guideline and with our findings in the spatial
analysis (Sect.~3.1, i.e., the radial profiles drop to zero around
2$^{\prime\prime}$), we consider a source extraction region of
2$^{\prime\prime}$ to contain all of the core flux.  Smaller
extraction regions (radius=1.5$^{\prime\prime}$) were used for two
objects (0802+103 and 0836+299) since the 2$^{\prime\prime}$ aperture
included emission from the X-ray jets. The background was estimated in
a nearby region free of obvious sources, with a radius of
8$^{\prime\prime}$. The net source counts in 0.5--8 keV after
background subtraction are reported in Table~2. We estimated the
total X-ray counts in the jet, by extracting the counts from a
rectangular region of varying widths and heights to best suit the
shape of the particular jet; the dimensions of the box are reported in
Table~2. Also listed in Table~2 is the ratio of the core-to-jet X-ray
counts (Col. 7); this ratio can be used to gauge the contribution of
the jet to the total X-ray emission in previous, lower-resolution
X-ray observations.

For the 10 sources affected by pileup of the nuclear flux, we list in
Table 2 the total counts collected in a 2$^{\prime\prime}$ radius
circle. Since pileup is known to affect the spectral index, we
extracted X-ray spectra for these objects using the counts in the
wings of the PSF.  For relatively low pileup rates, pileup is not
believed to affect the wings of the PSF, only its core (within
0.5$^{\prime\prime}$). We extracted the spectrum using an annulus
centered on the source centroid with inner radius R$_{in}$ and outer
radius fixed at R$_{out}=3^{\prime\prime}$.  We experimented with
various choices of R$_{in}$ (from 0.5$^{\prime\prime}$\ to
1.5$^{\prime\prime}$), and selected the radius that gave the maximum
number of counts for spectral analysis, R$_{in}$=0.5$^{\prime\prime}$.
The ancillary files were corrected for the chosen annular extraction
region, and used for the spectral fits.

\subsubsection{Spatial Analysis}

Spatial analysis was performed on all 17 sources. Our aim was to
investigate whether diffuse X-ray emission is present around the
cores, or whether the X-ray radial profile is entirely consistent with
the instrumental point spread function (PSF).

The instrumental PSFs were created using the PSF library for the ACIS
camera at 1.5 keV, from the {\it Chandra} Calibration Database
(CALDB).  The CIAO tool \verb+mkpsf+ simulates a point source, and the
tool \verb+dmimgcalc+ normalizes the instrumental PSF to the count
rate of a specific nucleus. The extraction of radial profiles
allows the data from this normalized image to be compared to the
observations.

Radial profiles were extracted using the \verb+CIAO+ tools
\verb+dmextract+ and \verb+dmtcalc+, after restricting the images to
the energy range of {\it Chandra} sensitivity, 0.5--8 keV, and
cleaning them of features that would contaminate the radial
profiles. Specifically, serendipitous X-ray sources in the field (as
for 0605$-$085, 1136$-$135) were removed, as well as X-ray
counterparts of extended radio features (Table 2, Col. 6).
Additionally, for piledup sources (Table 2, Col. 5), the charge
transfer trail from the nucleus was removed.  The \verb+CIAO+
tools \verb+dmextract+ and \verb+dmtcalc+ only calculate a
monochromatic PSF, while it is well known that the ACIS PSF is
energy-dependent.  Following Donato et al. (2003) and Worrall et
al. (2001), we calculated PSF profiles at eight different energies
(0.73, 0.98, 1.23, 1.63, 2.13, 3.03, 5.48, 7.23 keV) and assigned a
weight according to the distribution of photons as a function of
energy. The PSF used for comparison with the observed radial
profiles is the weighted sum of the energy-dependent PSFs.

Upon the cleaned images, centered on the source centroid, forty annuli
of equal width (0.5$^{\prime\prime}$) were placed in order to extract
the radial profiles out to 20$^{\prime\prime}$. The background was
extracted from an annular region, centered on the source centroid with
an inner radius R$_{in}$=21$^{\prime\prime}$\ and an outer radius
R$_{out}$=26$^{\prime\prime}$, outside the region of the profile. For
the nearby FRI radio galaxy 0836+299, the radial profiles were
extracted from wider annuli (1.25$^{\prime\prime}$ radius) up to
25$^{\prime\prime}$ from the core.  The background for this object was
extracted from an annular region with an inner radius
R$_{in}$=26$^{\prime\prime}$\ and an outer radius
R$_{out}$=31$^{\prime\prime}$.  This is the maximum distance allowed
in the 1/8 subarray mode.  For sources affected by core pileup we
normalized the PSF to the observed radial profile at the fourth pixel,
(as suggested by the {\it Chandra} POG) otherwise, the final PSF was
normalized to the first pixel.

\subsubsection{Timing Analysis}

To check for time variability of the source flux within the {\it
Chandra} exposure, we extracted light curves of the cores using the
\verb+FTOOL+ ``lcurve'' v. 1.0 (XRONOS 5.16).  Light curves were
extracted from the 2$^{\prime\prime}$ radius core region in the 0.5--8
keV energy range. The light curves were plotted at a variety of
binnings (100--750 second intervals) to check for patterns that might
suggest short-timescale variability. To decide if variability was
present, we performed a standard $\chi^{2}$ test. We tested for
significant deviations from the average count rate within the light
curves.  The value for $\chi^{2}$ varied with the different binnings
but not significantly.  Listed in Table 3 are the $\chi^2$ values
calculated from the 500 second bins and the probability P$_{\chi^2}$
at this binning.  This is the probability that the light curve is
constant.

\subsubsection{Spectral Analysis}

Spectral analysis was performed within \verb+XSPEC+ v.11.0.1. To
account for the recently observed quantum efficiency decay of ACIS,
possibly caused by molecular contamination of the ACIS filters, we
have applied a time-dependent correction to the ACIS quantum
efficiency based on the presently available information from the
CXC. The fits were restricted to the energy range 0.5--8.0~keV, where
the background contribution is negligible and where the ACIS-S
calibration is best known, especially regarding the ACIS-S gain.

Generally, we performed spectral analysis only for sources where 100
counts or more were detected from the core (unpiled sources) or from
the PSF wings (piledup sources; see above).  The spectra were rebinned
to have a minimum of 20 counts in each new bin, in order to apply the
$\chi^2$ statistics.  For the low-power source 0836+299, only 80
counts were detected from the core. Although this is formally less
than our threshold number of counts for spectral analysis, inspection
of the distribution of counts versus energy in this source reveals an
interesting spectrum (see below). In light of our interest in this
source (the only low-power, FRI source of our sample), we performed
spectral analysis for 0836+299 with a different binning (15 cts/bin).

To begin, we fitted all sources with a single power law model with an
absorption column density fixed to the Galactic value (Table 1,
Col. 5), using solar abundances and the Morrison \& McCammon (1983)
cross sections.  Next, several multi-component models were fitted to
the ACIS spectra of the nuclei, until the minimum $\chi^2$ was
obtained. The significance of adding more complex models to the fit
was estimated via the F-test probability over the original fit.  We
report both the simple power law fits and the significant
multi-component fits.

\section{Results}

Below we present the results from the spatial, timing, and spectral
analysis for the whole sample. Tables 3--7 report the results in
numerical form. 

\subsection{Spatial Properties}

We find that in 16/17 sources there is no evidence for diffuse X-ray
emission around the cores in the short {\it Chandra} exposures: the
background-subtracted radial profiles drop to zero around
2$^{\prime\prime}$, with most ($\sim$90\%) of the flux contained
within 1$^{\prime\prime}$. In these cases, the observed radial profile
is entirely consistent with the instrumental PSF. Figure 1a shows the
radial profile for 0405$-$123, which is an example of the 16
quasi-stellar sources without diffuse emission, together with the
residuals (ratios of the data to the PSF). 

\begin{figure}[h]
\resizebox{\hsize}{!}{\includegraphics{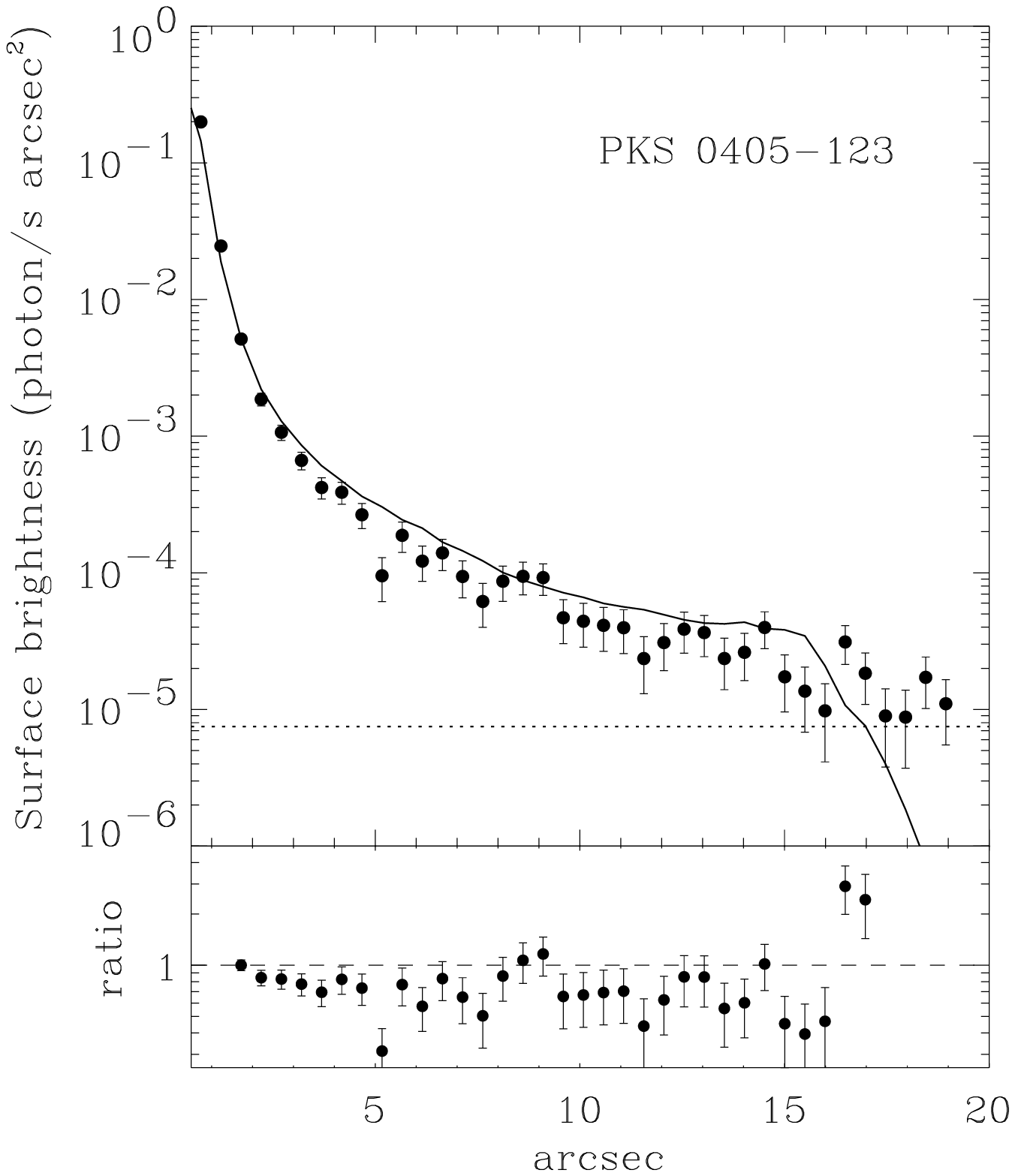}}\vspace{0.5cm}\\
\resizebox{\hsize}{!}{\includegraphics{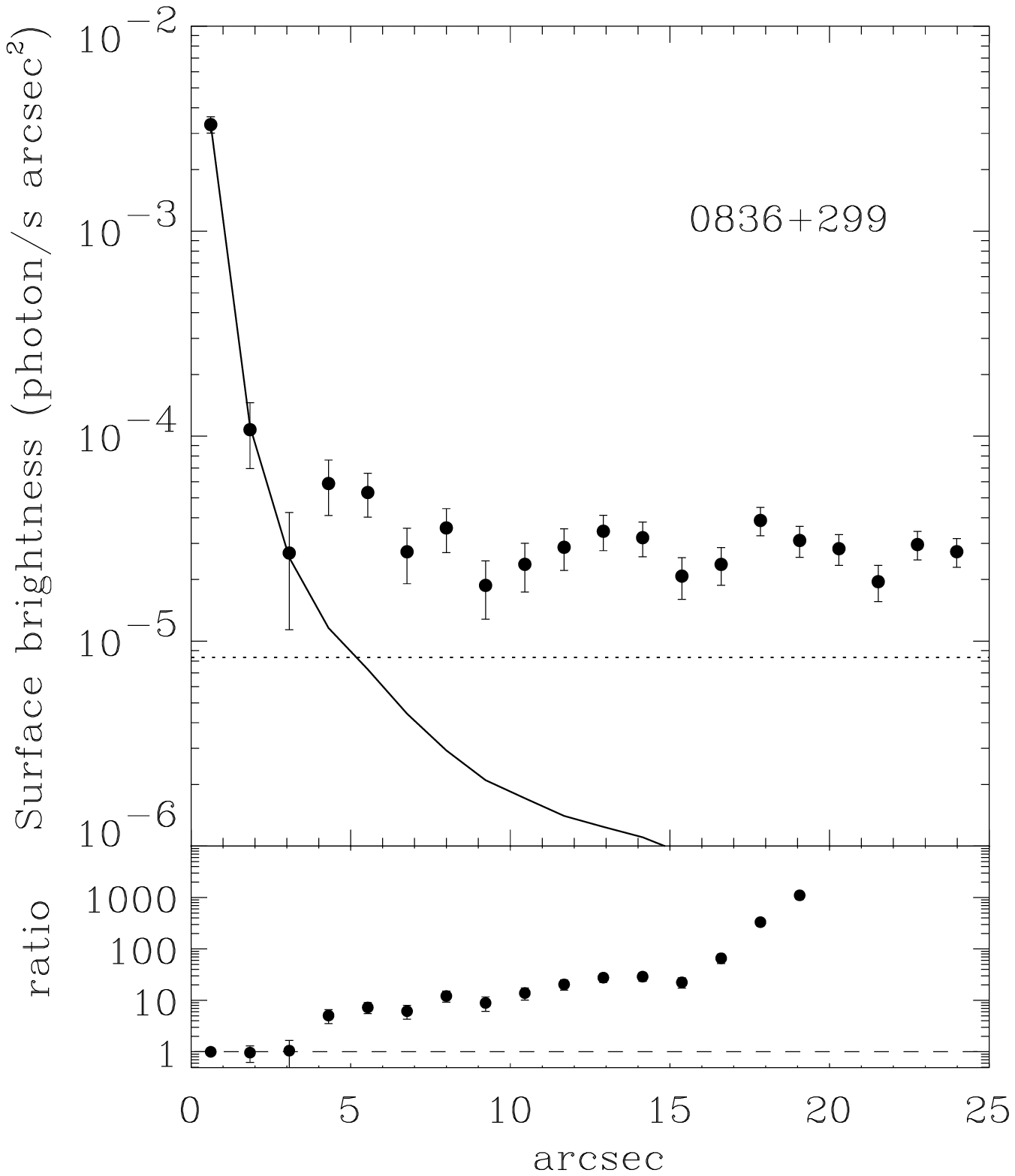}}\\
\caption{{\it (a, Top)} Radial profile for 0405$-$123.  The
solid line is the instrumental profile for a point source.  The dotted
line is the background level.  The plotted ratio of the surface
brightness to the point-spread function shows that there is no
extended emission present near this object, as there is none for the
quasars of our sample.  {\it (b, Bottom)} Radial profile for the
nearby, low-power radio galaxy 0836+299. In this case, there is
evidence for diffuse X-ray emission on scales of a few kpc, consistent
with other FRIs (Worrall et al. 2001).}
\end{figure}

However, evidence for diffuse X-ray emission is present in the
case of 0836+299, the only FRI of the sample. As Figure 1b clearly
shows, this object has an excess of X-ray emission over the PSF and
the background at 2--3$\sigma$ level at 5$^{\prime\prime}$ from the
core. This is in keeping with the results of Worrall et al. (2001),
who find that extended X-ray halos are a common property of low-power
radio galaxies.

We tried a fit to the radial profile with a King profile;
however, the parameters of the model (the $\beta$ value and the core
radius) are unconstrained. The intrinsic 0.4--5 keV luminosity of the
diffuse emission is L$_{0.4-5~keV}= 3.3\times 10^{41}$ ergs~s$^{-1}$
(assuming a Raymond-Smith thermal plasma with temperature $kT=1$ keV,
solar abundances, and a Galactic column density), within the range
measured for other FRIs (L$_{0.4-5~keV} \sim 2.6-4.2 \times 10^{41}$
ergs~s$^{-1}$ (Worrall et al. 2001). Deeper ACIS observations are
needed to confirm the presence of diffuse X-ray emission around the
core of 0836+299 and better quantify its properties.

We can determine limits for the detection of galaxy halos for all
of the quasars of the sample, since each has a core radial profile
that is consistent with a point source.  Any small-scale diffuse X-ray
emission, if present, is confined within 1$^{\prime\prime}$,
corresponding to 3.7 kpc at $z$=0.302 and 5.5 kpc at $z$=1.956, the
two extreme redshifts of the quasar sample. In this case, while we can
not resolve the diffuse component spatially, we could in principle
detect a thermal emission component in the X-ray spectrum of the core.
Alternatively, any diffuse larger-scale X-ray emission, if present on
scales $>$1$^{\prime\prime}$, is too weak to be detected in our short
{\it Chandra} exposures, and we can determine a detection threshold.

To assess our sensitivity to the thermal component, we performed a
simulation with \verb+XSPEC+ assuming a power law model (associated
with the active nucleus) with photon index $\Gamma=$1.66 (average
value from the sample) and flux F$_{2-10~keV}=8 \times 10^{-13}$
ergs~cm$^{-2}$~s$^{-1}$, at a median redshift of $z$=0.6. We added a
Raymond-Smith thermal component with temperature $kT=1$ keV (typical
of FRIs' halos), abundance fixed to 0.5 solar, and varied its
normalization until the thermal component was no longer detected
significantly, according to the F-test. We find that the latter occurs
for a flux, F$_{0.4-5~keV}=5.9 \times 10^{-14}$
ergs~cm$^{-2}$~s$^{-1}$, which corresponds to an intrinsic luminosity,
L$_{0.4-5~keV}=4.8 \times 10^{43}$ ergs~s$^{-1}$, greater than the
typical halo luminosity measured for FRIs.  A similar flux limit is
obtained for 0838+133, although it has a longer exposure, since its
X-ray core is 4 times brighter than our average model.

To assess our sensitivity to diffuse emission on larger-scales
($>$1$^{\prime\prime}$, $\sim$10 kpc), we performed \verb+MARX+
simulations. We assumed a point source with an average core X-ray flux
of $8 \times 10^{-13}$ ergs~cm$^{-2}$~s$^{-1}$, at a median redshift of
$z$=0.6 and a spectrum described by a power law with a photon index of
1.66. Following Worrall et al. (2001), the point source was described
spatially as a Gaussian with width $\sigma=0.2$ to account for the
aspect solution. For the diffuse X-ray emission we assumed a modified
King profile for cluster emission, with the core radius fixed at
2$^{\prime\prime}$ (or 10 kpc intrinsic size for $z$=0.6), and the
$\beta$ parameter fixed at 0.8, and varied the 0.5--8 keV flux. We
find that diffuse X-ray emission can be detected down to a surface
brightness of $1 \times 10^{-14}$ ergs cm$^{-2}$ s$^{-1}$ arcsec$^{-2}$
for a 10 ks exposure, corresponding to the luminosity,
L$_{0.4-5~keV}=8.5\times 10^{43}$ ergs~s$^{-1}$.

We conclude that, with the exception of the nearby low-power galaxy
0836+299, no diffuse X-ray emission related to the host galaxy's halo
or cluster was detected around the cores of the quasars in our sample,
even in the case of 0838+133, which was observed for a significantly
longer exposure (37 ks).  Since our exposures were short ($\sim$10 ks)
and the detected nuclear emission is relatively strong, it is possible
that the diffuse emission is just too weak to be detected against a
very bright core in these relatively distant quasars.  Our simulations
show that any unresolved ($\;\buildrel<\over\sim\;$1$^{\prime\prime}$)
diffuse emission on the scale of the galaxy halo would be fainter than
F$_{0.4-5~keV}=5.9 \times 10^{-14}$ ergs~cm$^{-2}$~s$^{-1}$, while any
larger-scale diffuse emission would have a surface brightness $\;
\buildrel < \over \sim \;$ $1 \times 10^{-14}$ erg cm$^{-2}$ s$^{-1}$
arcsec$^{-2}$.

\subsection{X-ray variability on short and long timescales}

Most (13/17) of the sources in this sample are classified as variable
at optical and other wavelengths, and two of them are also known from
previous studies to exhibit long- and short-term variations of the
X-ray flux.  The blazar 1510$-$089 showed a factor of $\sim$3.5
variation in flux on a timescale $\sim$2 years, between survey and
pointed observations with {\it ROSAT} (Siebert et al. 1996).  The
blazar 1641+399 showed a factor of $\sim$2 variation also on a
timescale $\sim$2 years (Sambruna 1997).  Here we examine whether
short-term variability is also present within our {\it Chandra}
exposures.

\begin{table}[b]
\begin{center}
\begin{tabular}{lccc}
\multicolumn{4}{l}{{\bf Table 3: Timing Analysis of Quasar Cores}} \\
\multicolumn{4}{l}{   } \\ \hline \hline
& & & \\
\multicolumn{1}{c}{Source} & $\chi^{2}_{r}$/dof & P$_{\chi^{2}}$  & Variable \\ 
\multicolumn{1}{c}{(1)} & (2) & (3) & (4) \\ 
& & & \\ \hline
& & &  \\
0405$-$123 & 1.064/19 & 0.382 & N \\
0605$-$085 & 0.261/19 & 0.999 & N \\
0723+679 & 0.881/20 & 0.612 & N \\
0802+103 & 0.458/18 & 0.975 & N \\
0836+299 & 0.436/17 & 0.976 & N \\ 
0838+133 & 0.650/24 & 0.183 & N \\
1040+123 & 0.999/23 & 0.462 & N \\
1055+018 & 0.534/20 & 0.954 & N \\ 
1136$-$135 & 0.676/19 & 0.846 & N \\
1150+497 & 0.863/20 & 0.636 & N \\
1354+195 & 0.638/19 & 0.880 & N \\
1510$-$089 & 2.626/21 & 3.6e-9 & Y \\ 
1641+399 & 1.005/19 & 0.451 & N \\
1642+690 & 0.771/18 & 0.737 & N \\
1741+279 & 0.440/19 & 0.982 & N \\
1928+738 & 1.405/18 & 0.115 & N \\
2251+134 & 0.570/20 & 0.935 & N \\
& & & \\ \hline 
\end{tabular}
\end{center}
{\bf Columns Explanation:} 1=Source IAU name; 2=Reduced
$\chi^{2}$ and degrees of freedom (dof) for 500 second binning;
3=Probability of $\chi^{2}$ statistics; 4=Flag indicating whether the
source is variable (Y), or not (N) based on column (3).
\end{table}

Table~3 reports the results of the $\chi^2$ test applied to the
500s-bins {\it Chandra} light curves of the core, and the probability
that the light curve is constant ($\chi^2$ probability). No
high-amplitude variability is present based on this test.

Only 1510$-$089 has a large enough reduced $\chi^2$ value,
$\chi_r^{2}$=2.63, to indicate possible variability.  Inspection of
its light curve shows erratic excursions of the X-ray flux, with no
apparent regular trend at the 500s binning.  For this reason the
Kolmogorov-Smirnov (KS) test for a probability of constancy was
applied to this object at a variety of binnings.  The results of this
test show that the probability of constancy for this object is 6.7\%
at binnings of 750, 250 and 100 seconds, but is 10\% at a binning of
500s, the most erratic bin set. In other words, according to the KS
test the probability of variability in the light curve ranges between
90\% and 93.3\%. Figure 2 shows the light curve for this object binned
at 750s. A change of the flux by a factor of 1.4 within 25 minutes is
apparent.

\begin{figure}[h]
\resizebox{\hsize}{!}{\includegraphics[angle=-90]{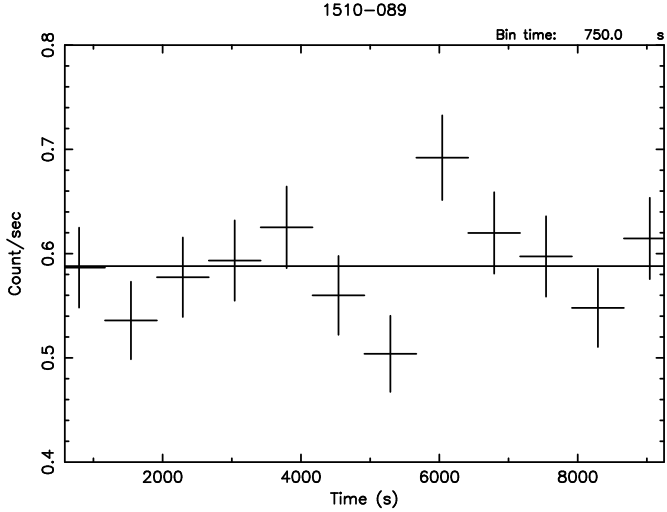}}
\caption{Light Curve of 1510$-$089 binned at 750s
intervals. Low-amplitude flux changes on a timescale of 25 minutes are
present.}
\end{figure}

We also investigated whether the flux variability in 1510$-$089 is
energy-dependent. Light curves binned at 750 s were accumulated in a
soft (0.3--1 keV) and a hard (2--8 keV) energy band, and the $\chi^2$
and KS tests were applied. At soft X-rays we find a 92\% probability
of variation in the light curve, and at hard X-rays we find a 96\%
probability of variation.  We conclude that variability is present at
both soft and hard X-rays in 1510$-$089.

In conclusion, none of the objects decisively demonstrated short-term,
high-amplitude variability in our {\it Chandra} exposures. The only
exception is 1510$-$089 which showed a low-amplitude flux change within
a timescale of $\sim$ 25 minutes.  We suggest long-exposure
observations with {\it Chandra} or {\it XMM-Newton} of this object for
confirmation of this result.

We also checked for long-term X-ray variability by comparing the {\it
Chandra} flux with previously published X-ray fluxes from other
experiments.  We decided to compare the monochromatic 1 keV fluxes,
because these are less sensitive to the shape of the X-ray continuum
and are not based on the different waveband sensitivities of earlier
experiments.  The {\it Chandra} monochromatic flux at 1~keV is
reported in Table~7 for each object, while previous X-ray observations
of the sources in our sample are listed in Appendix A.  As a
caveat it is worth noting that the X-ray fluxes from the pre-{\it
Chandra} experiments used much larger extraction radii (of the order
of arcminutes); however, as shown by our spatial analysis, the nucleus
is the dominant source of X-ray emission in these regions.  In
Table~7, five objects are found to be long-term X-ray variables:
0405$-$123 shows a factor of $\sim$1.5 decrease in 9 years (Sambruna
1997); 1055+018 a factor of $\sim$1.5 increase in 8 years (Siebert et
al. 1998); 1354+195 a factor of $\sim$3 decrease in $\sim$20 years
(Biermann et al. 1987); 1510$-$089 a factor of $\sim$1.5 decrease in
$\sim$8 years (Siebert et al. 1998 and Singh et al. 1990); and
1928+738 a factor of $\sim$2 decrease in 10 years (Lawson et
al. 1992).

\subsection{X-ray continuum emission}

\noindent{\bf Quasar cores:} The results of the spectral fits with a
single power law plus fixed Galactic absorption are reported in Table
4, including the observed 2--10 keV fluxes, and the intrinsic
(absorption-corrected) 2--10 keV luminosities.  All fitted parameters
are listed with their 90\% confidence errors, ($\Delta \chi^{2}$=2.7
for one parameter of interest).  

\begin{table}[h]
\begin{center}
\begin{tabular}{rlcccc}
\multicolumn{6}{l}{{\bf Table 4: Single Power Law Fits for the X-ray Continua}$^a$} \\
\multicolumn{6}{l}{   } \\ \hline \hline
& & & & &  \\
\multicolumn{1}{c}{Source} & \multicolumn{1}{c}{$\Gamma$} & $\chi^{2}_{r}/dof$ &F$_{2-10\ keV}$ & L$_{2-10\ keV}$\\ 
\multicolumn{1}{c}{(1)} & \multicolumn{1}{c}{(2)} & (3) & (4) & (5) \\
& & & & & \\ \hline
\\
0405$-$123 & $1.90\pm 0.05$ & 1.32/152 & 4.8$\times 10^{-12}$& 2.7$\times 10^{45}$\\
0605$-$085 & $1.66 \pm 0.09$ & 1.03/55 &  1.1$\times 10^{-12}$ & 1.2$\times 10^{45}$ \\
0723+679 & $1.76 \pm 0.08$ & 0.99/71 & 9.1$\times 10^{-13}$ & 9.4$\times 10^{44}$  \\
0802+103 & $1.88 \pm 0.23$ & 0.60/9 & 1.2$\times 10^{-13}$ & 4.4$\times 10^{44}$ \\
0838+133 & $1.35 \pm 0.06$ & 1.25/104 & 1.2$\times 10^{-12}$ & 7.6$\times 10^{44}$\\
1040+123 & $1.69 \pm 0.07$ & 1.26/76 & 9.7$\times 10^{-13}$ & 1.3$\times 10^{45}$ \\
1055+018 & $1.67 \pm 0.07$ & 1.12/86 & 2.7$\times 10^{-12}$ & 3.0$\times 10^{45}$\\
1136$-$135 & $1.76\pm 0.09$ & 0.99/44 &  1.2$\times 10^{-12}$ & 1.8$\times 10^{45}$ \\
1150+497 & $1.86 \pm 0.06$ & 1.22/86 & 2.2$\times 10^{-12}$ & 4.3$\times 10^{44}$\\
1354+195 & $1.56 \pm 0.07$ & 1.06/75 & 2.6$\times 10^{-12}$ & 6.4$\times 10^{45}$ \\
1510$-$089 & $1.40\pm 0.06$ & 1.31/102 & 5.1$\times 10^{-12}$ & 1.1$\times 10^{45}$\\
1641+399 & $1.71 \pm 0.07$ & 1.18/82 & 2.6$\times 10^{-12}$ & 1.5$\times 10^{45}$ \\
1642+690 & $1.77 \pm 0.10$ & 0.96/51 & 8.0$\times 10^{-13}$ & 6.8$\times 10^{44}$\\
1741+279 & $2.05\pm 0.13$ & 0.98/36 & 8.0$\times 10^{-13}$ & 2.0$\times 10^{44}$ \\
1928+738 & $1.88\pm 0.07$ & 1.01/98 & 3.3$\times 10^{-12}$ & 5.4$\times 10^{44}$ \\
2251+134 & $1.82 \pm 0.08$ & 1.13/65 & 7.8$\times 10^{-13}$ & 5.6$\times 10^{44}$ \\
& & & &\\ \hline
\end{tabular}
\end{center}
{\bf Columns Explanation:} {1=Source IAU name; 2=Photon index
and 90\% uncertainties ($\Delta\chi^2=2.7$); 3=Reduced $\chi^2$ value
and degrees of freedom (dof); 4=Absorbed flux in 2--10~keV in ergs cm$^{-2}$
s$^{-1}$ ; 5=Intrinsic X-ray luminosity in 2-10~keV assuming redshift
(Table~1) and cosmology ($H_{0}$~ =~75 $\frac{km/s}{Mpc},
q_{0}~=~0.5$) in ergs s$^{-1}$.}

\noindent{\bf Notes:} {$a$=Radio galaxy 0836+299 did not have an
acceptable single power law fit (Sect.~3.3) and is listed in Table~5.}
\end{table}
\begin{table*}[]
\begin{center}
\begin{tabular}{rlcccc}
\multicolumn{6}{l}{{\bf Table 5: More Complex Continuum Fits}} \\
& & & & \\ \hline \hline
& & & & & \\
\multicolumn{1}{c}{Source} & \multicolumn{1}{c}{Best-Fit Model and Parameters} & $\chi^{2}_{r}/dof$  & $\Delta \chi^{2}$/P$_{F}$ & F$_{2-10\ keV}$ & L$_{2-10\ keV}$\\ 
\multicolumn{1}{c}{(1)} & \multicolumn{1}{c}{(2)} & (3) & (4) & (5) & (6) \\ 
& & & &  \\ \hline
\\
0405$-$123 & $\Gamma_{SOFT}=2.37^{+0.18}_{-0.15}$ & 1.14/150 & 31.7/99\% & 5.4$\times 10^{-12}$ & 2.9$\times 10^{45}$ \\
& (E$_0$=$1.07^{+0.13}_{-0.11}$keV,$\Gamma_{HARD}=1.74^{+0.07}_{-0.08} $)\\
0836+299 & power law ($\Gamma=1.69$, fixed) and  & 0.83/7 & $\cdots$ & 1.1$\times 10^{-12}$ & 3.4$\times 10^{43}$ \\ & absorption (N$_H$=$4.8^{+1.9}_{-1.4} \times 10^{23}$ cm$^{-2}$) & & & \\
1055+018 & $\Gamma_{SOFT}=2.11^{+0.27}_{-0.26}$ & 1.02/84 & 11.4/99\% & 3.0$\times 10^{-12}$ & 3.1$\times 10^{45}$ \\
& (E$_0$=$1.05^{+0.34}_{-0.17}$keV,$\Gamma_{HARD}=1.54^{+0.08}_{-0.09}$)\\
1150+497 & $\Gamma_{SOFT}=2.14^{+0.16}_{-0.13}$ & 1.07/83 & 16.3/99\% & 2.4$\times 10^{-12}$ & 4.6$\times 10^{44}$ \\
& (E$_0$=$1.37^{+0.60}_{-0.25}$keV,$\Gamma_{HARD}=1.74^{+0.11}_{-0.12}$)\\
1354+195 & $\Gamma_{SOFT}=1.88^{+0.37}_{-0.28}$ & 0.83/73 & 19.4/99\%  & 2.9$\times 10^{-12}$ & 6.8$\times 10^{45}$ \\
& (E$_0$=$1.09^{+0.32}_{-0.20}$keV,$\Gamma_{HARD}=1.39^{+0.10}_{-0.09} $)\\
1510$-$089 & $\Gamma_{SOFT}=1.68^{+0.24}_{-0.16}$ & 1.21/100 & 12.3/99\% & 5.7$\times 10^{-12}$ & 1.2$\times 10^{45}$ \\
& (E$_0$=$1.57^{+0.53}_{-0.44}$keV,$\Gamma_{HARD}=1.22^{+0.16}_{-0.12}$)\\
1641+399 & $\Gamma_{SOFT}=2.05^{+0.14}_{-0.13}$ & 0.88/80 & 26.6/99\% & 3.3$\times 10^{-12}$ & 1.7$\times 10^{45}$ \\
& (E$_0$=$1.59^{+0.31}_{-0.22}$keV,$\Gamma_{HARD}=1.37^{+0.12}_{-0.15}$)\\
& & & &\\ \hline
\end{tabular}
\end{center}
{\bf Columns Explanation:} {1=Source IAU name; 2=Best-fit
models: for all objects, except 0836+299, a broken power law is
used. In this model, E$_0$ is the break energy, $\Gamma_{SOFT}$ is the
photon index below E$_0$, $\Gamma_{HARD}$ is the photon index above
E$_0$; 3=Reduced $\chi^2$ and degrees of freedom (dof); 4=$\Delta
\chi^{2}$ from the single power law model and P$_F$ is the F-test
probability; 5=Absorbed flux in 2-10~keV in ergs~cm$^{-2}$~s$^{-1}$;
6=Intrinsic X-ray luminosity in 2-10~keV assuming redshift (Table~1)
and cosmology ($H_{0}$~ =~75 $\frac{km/s}{Mpc}, q_{0}~=~0.5$) in
ergs~s$^{-1}$.}
\end{table*}

In general, a single power law plus Galactic absorption provides an
acceptable description of the 0.5--8 keV continua, as judged by
inspection of the residuals and from the values of $\chi_r^2$ from
Table 4.  For two sources (0838+133 and 1040+123), this model provides
a large $\chi_r^2 \sim$ 1.26; inspection of the residuals shows the
presence of random fluctuations indicating a noisy spectrum.  Complex
models do not improve the fits of these two objects.

The single power law plus Galactic N$_H$ model is not an adequate
description of the X-ray continuum in 6 quasars (0405$-$123, 1055+018,
1150+497, 1354+195, 1510$-$089, and 1641+399). These sources show
complex residuals which require additional spectral components at low
energies. In all six cases, excess flux at soft X-rays ($\sim$1~keV)
is present over the extrapolation of the power law model. We modeled
the soft excess flux in terms of three possible alternative models: a)
thermal bremmstrahlung, b) blackbody, and c) broken power law. In all
cases, a broken power law with a steep index below $\sim$1.6 keV is
the preferred model based on the $\chi^2_r$. These best-fit models are
reported in Table 5.

No excess absorption was detected in the quasars of our sample,
including the quasar 0838+133, for which Brunetti et al. (2002) found
excess column density over Galactic. After making a correction for
the molecular contamination of ACIS, the column density we measured in
0838+133 was consistent with Galactic values.

\noindent{\bf The low-power radio galaxy 0836+299:} In this source,
inspection of the spectrum shows a marked deficit of counts below
2~keV, indicating large excess absorption over the Galactic value. A
formal fit with a single power law plus free absorption converges to a
negative value of the photon index and zero column density, which
simply parameterizes excess absorption at low energies, and is not
physically acceptable. We fitted the spectrum of 0836+299 with a power
law plus Galactic N$_H$ model, adding an extra column density at the
redshift of the source in the assumption that the excess absorption is
intrinsic (this is a nearby source). Because of the limited number of
counts, the photon index of the power law was fixed to the average
found for the other lobe-dominated sources ($\Gamma=1.69$; see below).
We found a significant intrinsic column density, N$_H \sim 5 \times
10^{23}$ cm$^{-2}$ (Table~5). The fitted column does not vary
significantly if the photon index is fixed to the extrema of the range
observed for this sample.

\subsection{Fe K-shell emission}

As discussed above (Sect.~1), Fe K-shell emission in higher-luminosity
sources, both radio-quiet and radio-loud, is not well
established. Interestingly, we find evidence for possible Fe K line
emission in two sources of our sample, 0723+679 and 1150+497, and
possibly in a third object, 1642+690. 

\begin{table}[]
\begin{center}
\begin{tabular}{ccccc}
\multicolumn{5}{l}{\bf Table 6: Fe K-shell Emission$^{a}$} \\
\multicolumn{5}{l}{ } \\ \hline \hline & & & & \\ 
\multicolumn{1}{c}{Source} & $E_{\ell}$ & EW & $\Delta$ $\chi^{2}$ & $P_{F}$ \\ (1) & (2) & (3) & (4) & (5) \\ & & & & \\\hline  
& & & & \\
\multicolumn{5}{c}{\bf Detections}\\ 
&&&&\\
0723+679 & $6.4^{+0.1}_{-0.1}$ & $164^{+455}_{-138}$ & 6.1 & 97\%\\ 
1150+497 & $6.4^{+0.6}_{-0.1}$ & $296^{+538}_{-203}$ & 5.0 & 97\% \\
1642+690 & $6.4^{+0.8}_{-0.7}$ & $188^{+748}_{-54}$ & 3.4 & 94\% \\
& & & & \\
\multicolumn{5}{c}{\bf Upper Limits$^{b,c}$}\\ 
& & & & \\
0405$-$123 & 6.4 & 294 & $\cdots$ & $\cdots$\\
0605$-$085 & 6.4 & 484 & $\cdots$ & $\cdots$\\
0802+103 & 6.4 & 865 & $\cdots$ & $\cdots$\\
0838+133 & 6.4 & 953 & $\cdots$ & $\cdots$\\
1040+123 & 6.4 & 528 & $\cdots$ & $\cdots$\\
1055+018 & 6.4 & 324 & $\cdots$ & $\cdots$\\
1136$-$135 & 6.4 & 992 & $\cdots$ & $\cdots$\\
1354+195 & 6.4 & 504 & $\cdots$ & $\cdots$\\
1510$-$089 & 6.4 & 434 & $\cdots$ & $\cdots$\\
1641+399 & 6.4 & 493 & $\cdots$ & $\cdots$\\
1741+279 & 6.4 & 908 & $\cdots$ & $\cdots$\\
1928+738 & 6.4 & 829 & $\cdots$ & $\cdots$\\
2251+134 & 6.4 & 751 & $\cdots$ & $\cdots$\\
& & & & \\ \hline
\end{tabular}
\end{center}
{\bf Columns Explanation:} {1=Source IAU name; 2=Rest-frame
line energy in keV, with line width fixed at $\sigma=0.05$ keV;
3=Rest-frame Equivalent width (EW) in eV; 4=$\Delta \chi^2$ from the
best-fit model (Table 4 or 5); 5=F-test probability.}

\noindent{\bf Notes:} {$a$=In all cases, an unresolved line
was assumed and the Gaussian width was fixed to $\sigma=0.05$~keV,
reported errors are 90\% confidence errors; $b$=No upper limit is
listed for 0836+299 (see text Sect.~3.4); $c$=Line Energy is fixed at
$E_{\ell}$=6.4 keV.}
\end{table}

\begin{figure*}[]
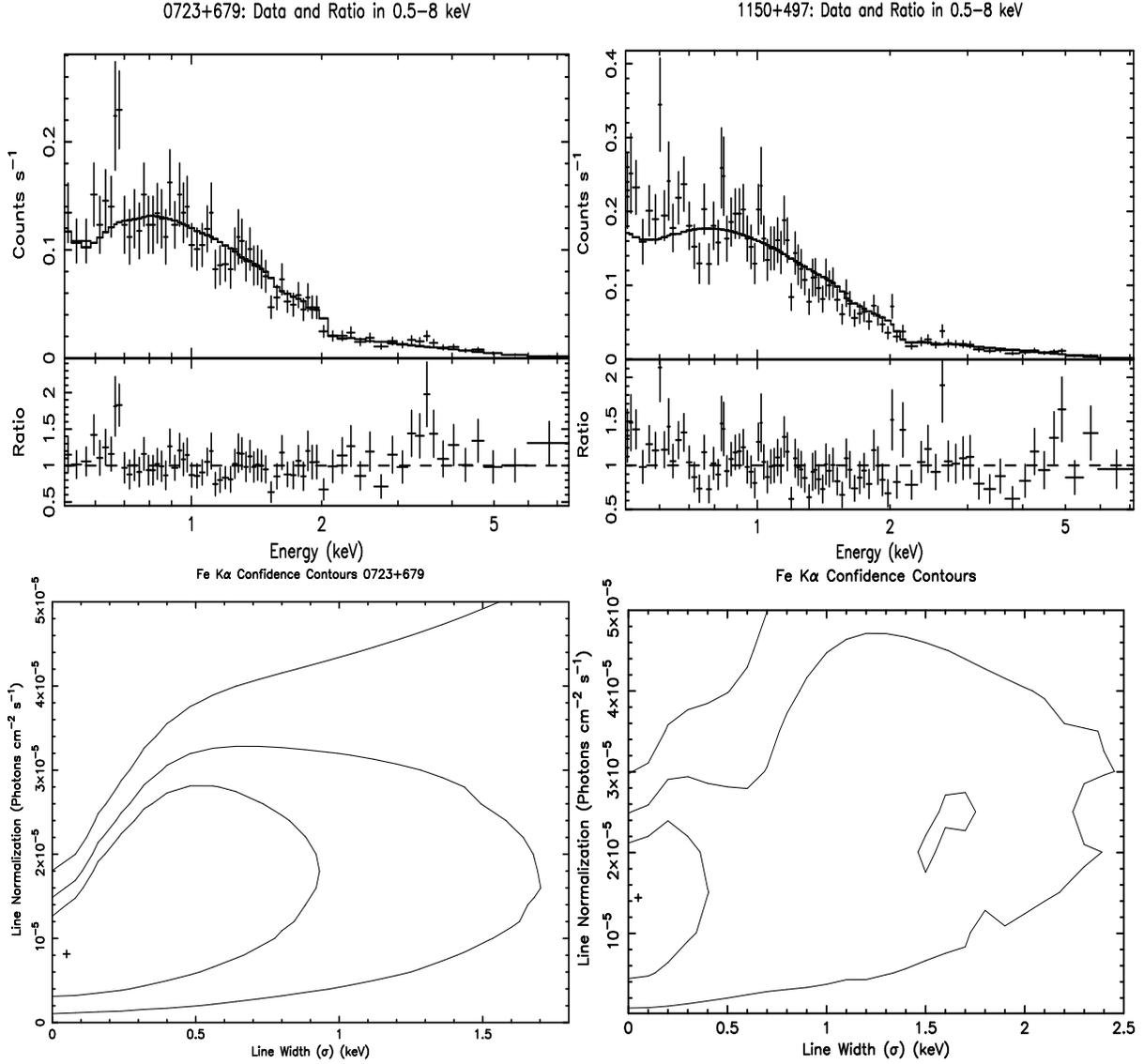

\begin{center}
{\includegraphics[width=8cm,height=8cm,angle=-90]{H4109F3a.ps}}{\includegraphics[width=8cm,height=8cm,angle=-90]{H4109F3b.ps}}
{\includegraphics[width=7cm,height=8cm,angle=-90]{H4109F3c.ps}}{\includegraphics[width=7cm,height=8cm,angle=-90]{H4109F3d.ps}}
\end{center}
\caption{Results of complex spectral fits of the quasars
0723+679 (left) and 1150+497 (right).  {\it Top Panels:} Observed
ACIS-S spectra and residuals of a single power law plus Galactic
absorption model. An Fe K line is present at observed energies 3.5 and
4.8 keV. {\it Bottom Panels:} Confidence contours at 68\%, 90\%, 99\%
confidence for the Gaussian line normalization versus width. In both
sources, the line is detected at ${\buildrel>\over\sim}$90\%
confidence and is always unresolved.}
\end{figure*}

Inspection of the residuals of the power law models shows excess
emission around 3.5 keV, 4.8 keV, and 3.7 keV for 0723+679, 1150+497,
and 1642+690, respectively. These energies are consistent with the
redshifted Fe K$\alpha$ line energy, 6.4 keV. To model the line, a
Gaussian component was added to the power law models reported in
Table~4 for each of the three sources. The significance of the
$\chi^2$ improvement is 97\% for 0723+679 and 1150+497, and 94\% for
1642+690. In Figure 3 we show the best-fit model and data for 0723+679
and 1150+497, for which the line detections are more significant. Also
shown in Fig. 3 are the confidence contours for the line flux versus
line width in both sources. The Fe K$\alpha$ line is detected at the
$\;\buildrel>\over\sim\;$90\% confidence level and is always
unresolved.  Since the line is unresolved, we fixed the line width at
$\sigma=0.05$ keV, lower than the S3 resolution (from Fig. 6.6 from
the {\it Chandra} POG, $\sigma\sim 0.2$ keV at 6 keV), and performed
the fit--- leaving the Gaussian center energy and normalization free
to vary. The fitted parameters (energy and EW, both in the rest-frame)
for a narrow line are listed in Table 6a, with their 90\% confidence
errors.  The fitted rest-frame emission line energies are consistent
with the K$\alpha$ fluorescent line from cold Fe. The fitted,
rest-frame EWs are 164$^{+455}_{-138}$, 296$^{+538}_{-203}$, and
188$^{+748}_{-54}$ eV for 0723+679, 1150+497, and 1642+690,
respectively.  Deeper observations of these targets with {\it Chandra}
and/or {\it XMM-Newton} are needed to confirm the line and study its
profile.

For the remaining 14 sources, we determined an upper limit to the Fe
K$\alpha$ line EWs. We fitted the spectra adding a Gaussian component
at an energy consistent with the redshifted energy of the Fe K$\alpha$
line and assumed an unresolved line ($\sigma=0.05$ keV). The 90\%
confidence upper limits on the line EWs (in the source's rest-frame)
are reported in Table~6b.  For one source (0836+299) the data included
too few counts from the hard X-ray spectrum to determine an upper
limit for the Fe K$\alpha$ emission line.

\section{Summary and Discussion}

We presented the X-ray properties of the cores of 16 radio-loud
quasars and one low-power radio galaxy observed with {\it Chandra} in
short ($\sim$10 ks) exposures. The targets were selected for their
large-scale radio properties, so the {\it Chandra} observations were
optimized for the detection of radio jets at X-rays. Thus, the sample
is not statistical as far as the core and extended properties are
concerned.

The {\it Chandra} resolution allows us to separate the nuclear X-ray
emission from the extended (kpc-scale) jet for the first time. In
general, we find that the cores dominate the total X-ray emission from
the source, with a large ratio of core-to-jet counts (Table~2,
Col.~6). The only exceptions are the lobe-dominated sources 0802+103
and 0836+299, where the core X-ray emission is comparable to and
smaller than the jet, respectively. However, it is necessary to stress
that the 2$^{\prime\prime}$ aperture used for the core spectrum
extraction corresponds to a region of intrinsic size $\sim$10 kpc at a
median redshift of $z$=0.6 ($\sim$2 kpc in the near source
0836+299). So, the ``core'' X-ray emission measured with {\it Chandra}
is likely contaminated by the inner regions of the jet.  Indeed, the
ratios of the jet-to-counterjet flux from radio observations (Cheung
et al. 2003) indicate that most of the jets in these sources are
closely aligned to our line of sight. Worrall et al. (2001) also found
that the X-ray jet is a substantial fraction of the total X-ray output
in the B2 FRI sample.

\subsection{Nuclear X-ray Emission}

The X-ray continua can be described by a power law, with further
spectral complexities (cold absorption in excess of Galactic, soft
excess, Fe K line) in 9/17 sources. Using the photon indices from the
best-fit models in Table 4 and 5, we calculated the average slope of
the 2--10 keV continuum. In the cases with a soft excess (Table 5), we
used the photon index above the break energy. The average photon index
for the whole sample is $\langle \Gamma_{sample} \rangle = 1.66$ with
standard deviation $\sigma_{sample}=0.23$.

It is interesting to compare the average X-ray slope from our sample
($\langle \Gamma_{sample} \rangle$=1.66, $\sigma_{sample}=0.23$), with
the slope derived for radio sources (BLRGs and quasars) with jets that
are known to be oriented at large angles to the line of sight
(Hasenkopf et al. 2002; Eracleous \& Halpern 1998). These sources have
inclination angles larger than 15$^{\circ}$\ (Eracleous \& Halpern
1998), or are classical radio doubles where the line of sight is not
too close to the jet (Hasenkopf et al. 2002). From {\it ASCA}, {\it
RXTE}, and {\it BeppoSAX} observations of these sources, an average
photon index of $\langle \Gamma \rangle=1.83$ and dispersion 0.04 is
obtained in a range of intrinsic X-ray luminosities similar to our
sources. The average slope for the more misaligned sources is steeper
than for our sources (within large dispersions), supporting the idea
that contamination from the unresolved, inner parts of a jet is
present in our sample. Indeed, the average photon index for our
quasars is consistent with the average X-ray photon index measured
with {\it BeppoSAX} for a sample of gamma-ray bright blazars
(Tavecchio et al. 2000), where beaming is important. The average slope
for our sample is also flatter than for radio-quiet quasars, where
$\langle \Gamma_{RQQ} \rangle=1.89$ and dispersion 0.05 (Reeves \&
Turner 2000).

Previous studies of radio-loud quasars with {\it ROSAT} (Elvis et
al. 1994) indicated the presence of excess cold absorption over the
Galactic value in high-z sources. At odds with this result, we do not
find evidence for excess X-ray absorption in our two $ z > 1$ quasars,
0802+103 ($z$=1.96) and 1040+123 ($z$=1.03). Interestingly, we do find
evidence for excess X-ray absorption in the {\it closest} source of
the sample, 0836+299 (see Sect.~4.3). 

In 6/17 sources, excess X-ray flux at soft energies is detected. Our
best-fit parameterization is a broken power law, with a steep
component below 1 keV. Inspection of the Spectral Energy Distributions
of these soft excess quasars using data compiled by us shows that the
best sampled sources of this group (0405$-$123, 1354+195, 1510$-$089,
and 1641+399) exhibit a Blue Bump (BB) in the optical part of the
spectrum. It is likely that the soft X-ray excess detected in our
ACIS-S spectra is related to the high-energy tail of the BB, for which
a possible origin is thermal emission from an optically thick
accretion disk (e.g., Czerny \& Elvis 1987). This would indicate the
presence of a strong isotropic radiation field in these sources which
is important for models of their high-energy emission (Tavecchio et
al. 2000). Alternatively, the soft X-ray excess could be related to
non-thermal (synchrotron) emission from the inner jet. In this case,
one could expect to observe significant short-term flux variability as
a result of synchrotron radiative losses. The possible detection of
flux changes on timescales of $\sim$25 minutes in 1510$-$089, if
confirmed, would indicate the presence of non-thermal emission at soft
X-rays in this source. Future {\it XMM-Newton} observations are needed
to confirm and better study the soft X-ray excess in the quasars of
our sample.

We detect an Fe K$\alpha$ emission line in two sources (0723+679 and
1150+497) and possibly in a third (1642+690) source. The detection of
such a line is very interesting {\it di per se}, because the presence
of Fe K-shell emission from high-luminosity AGNs in general, and from
radio-loud sources in particular, is still not established (see
Sect.~1). While it is important to confirm these lines and study their
profile with higher-quality X-ray spectra, we note that their measured
rest-frame EWs (Table~6a) are consistent with the EWs of the other
radio-sources with a double-lobe radio morphology and similar
intrinsic X-ray luminosity recently studied by {\it ASCA} (Hasenkopf
et al. 2002). Usually, dilution of the X-ray continuum by a beamed jet
component would weaken the strength of the Fe K lines in radio-loud
quasars. This could suggest weaker beaming in 0723+679 (indeed a SSRQ)
and in 1150+497 and 1642+690 (both FSRQs).

Although beaming is present in all the quasars of our sample as
indicated by their broad-band X-ray continuum properties, we conclude
that beaming is likely to affect the individual sources to a different
degree.

\subsection{Core- and Lobe-Dominated Sources}

Splitting the sample into core- and lobe-dominated sources based on
Table 1 (Col. 7), we find no difference in the average photon index
for the two subgroups: for the 11 core-dominated quasars, $\langle
\Gamma_{cd} \rangle = 1.65$ and $\sigma_{cd}=0.25$, and for the five
lobe-dominated sources $\langle \Gamma_{ld} \rangle = 1.69$ with
$\sigma_{ld}=0.20$.  If we exclude the unusually flat source 0838+133
(the only lobe-dominated source with strong core pileup), the range of
the photon index for the lobe-dominated sources is greatly
constrained: $\langle \Gamma_{ld} \rangle = 1.77$ and
$\sigma_{ld}=0.08$.  

Figure 4 shows the plot of the X-ray photon index from Table 4 or 5
versus the ratio of core-to-extended flux $R_i$ from Table 1. Our
measurements of the extended radio flux included contributions from
the kpc-scale jets, which are presumably still beamed to some degree,
so our values of $R_i$ do not follow the usual convention (Orr \&
Browne 1982).  Nevertheless our derived values are consistent with
values from the literature, where available (Cheung et al. 2003) and
can be used to gauge the relative importance of beaming in these
objects.  There is neither any trend between the slope of the X-ray
photon index and $R_i$ for the whole sample, nor for individual
subgroups. The distribution of slopes for the core-dominated quasars
overlaps with the distribution for lobe-dominated ones, and extends to
flatter slopes; 4/11 core-dominated sources have slopes $\Gamma$ {$\;
\buildrel < \over \sim \;$} 1.6, while 4/5 lobe-dominated are steeper
than this value.

\begin{figure}[h]
\resizebox{\hsize}{!}{\includegraphics[angle=0]{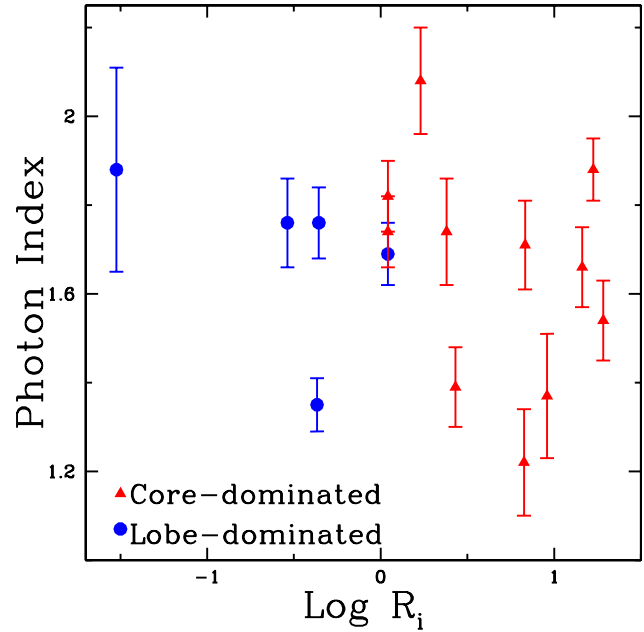}}
\caption{Plot of the best-fit photon index of the X-ray continuum
versus the core-dominance parameter $R_i$ for the 16 quasar
sample. There is no trend for the whole quasar sample, nor within the
subgroups of core- and lobe-dominated sources. The measured X-ray
continuum slopes are similar for both subgroups.}
\end{figure}

We conclude that no difference in X-ray spectral slope is present
between the core- and lobe-dominated subgroups of our sample. This
result is not unexpected, as all the quasars including the SSRQs have
one-sided radio jets and beaming is important in all cases.  Larger
samples of both core- and lobe-dominated quasars with optimal
selection criteria are needed to verify these X-ray spectral
properties.

\begin{figure}[]
\resizebox{\hsize}{!}{\includegraphics[angle=0]{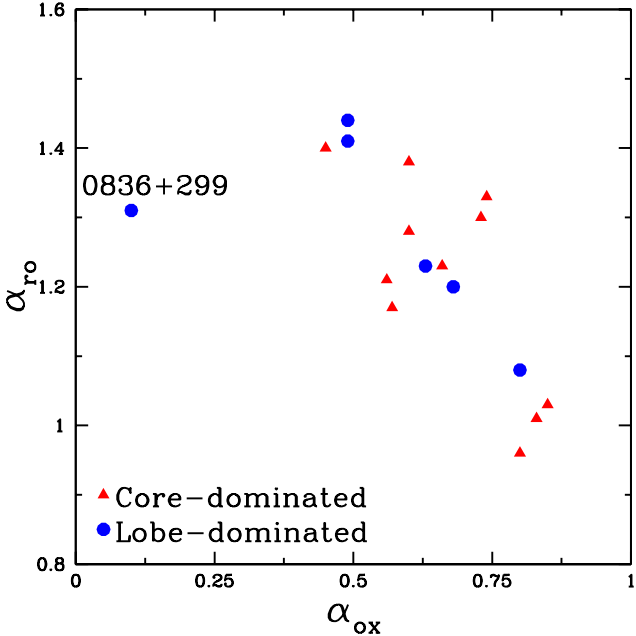}}
\caption{Broad-band spectral properties of the quasars and of the
radio galaxy of our sample.  There is no difference between the core-
and lobe-dominated subgroups. The FRI galaxy 0836+299 sticks out for
its flat optical-to-X-ray flux ratio.}
\end{figure}

We also compared the broad-band spectral distributions of both quasar
subclasses. Broad-band spectral indices between radio (5 GHz) and
optical (V band), $\alpha_{ro}$, between optical and X-rays (1 keV),
$\alpha_{ox}$, and between radio and X-rays, $\alpha_{rx}$, were
calculated using published values and the {\it Chandra} fluxes from
this work. Figure 5 shows the plot of the broad-band indices,
$\alpha_{ro}$ versus $\alpha_{ox}$, while the broad-band monochromatic
values and the plotted indices are listed in Table 7. We see no marked
difference between core- and lobe-dominated objects from the
broad-band spectra. Overall, the quasars of our sample occupy a region
of steeper radio-to-optical flux ratios than usually found for blazars
(e.g., Donato et al. 2001), possibly as a result of their selection
criteria based on prominent radio emission.  The outlier in Figure 5
is the FRI source 0836+299 (see below), which stands out for its
flatter $\alpha_{ox}$ than the rest of the sample.

\begin{table}[]
\begin{center}
\begin{tabular}{lccclll}
\multicolumn{7}{l}{{\bf Table 7: Core Broad-Band Emission}} \\
\multicolumn{7}{l}{   } \\ \hline \hline
& & & \\
\multicolumn{1}{c}{Source} & \multicolumn{1}{c}{F{\small$_{5\ GHz}$}} & F{\small$_{5852\ \AA}$} & F{\small$_{1\ keV}$} &\multicolumn{1}{c}{$\alpha_{ro}$} & \multicolumn{1}{c}{$\alpha_{ox}$} &\multicolumn{1}{c}{$\alpha_{rx}$} \\
\multicolumn{1}{c}{(1)} & (2) & (3) & (4) & \multicolumn{1}{c}{(5)} & (6) & (7) \\
& & & \\ \hline
& & & \\
0405$-$123 & 0.961 & 5.6 & 1.01 & 0.45 & 1.40 & 0.78 \\
0605$-$085 & 2.252 & 0.50 & 0.17 & 0.73 & 1.30 & 0.93 \\
0723+679 & 0.423 & 0.31 & 0.16 & 0.63 & 1.23 & 0.84 \\
0802+103 & 0.040 & 0.18 & 0.03 & 0.47 & 1.41 & 0.80 \\
0836+299 & 0.008 & 2.48 & 0.80 & 0.10 & 1.31 & 0.52 \\
0838+133 & 0.483 & 0.19 & 0.12 & 0.68 & 1.20 & 0.86 \\
1040+123 & 1.163 & 0.12 & 0.16 & 0.80 & 1.08 & 0.89 \\
1055+018 & 3.256 & 0.22 & 0.44 & 0.83 & 1.01 & 0.89 \\
1136$-$135 & 0.463 & 1.67 & 0.23 & 0.49 & 1.44 & 0.82 \\
1150+497 & 0.443 & 0.61 & 0.47 & 0.57 & 1.17 & 0.78 \\
1354+195 & 1.755 & 1.69 & 0.34 & 0.60 & 1.38 & 0.87 \\
1510$-$089 & 1.482 & 1.48 & 0.56 & 0.60 & 1.28 & 0.84 \\
1641+399 & 8.133 & 1.59 & 0.44 & 0.74 & 1.33 & 0.95 \\
1642+690 & 1.403 & 0.08 & 0.14 & 0.85 & 1.03 & 0.91 \\
1741+279 & 0.204 & 0.31 & 0.18 & 0.56 & 1.21 & 0.79  \\
1928+738 & 2.991 & 1.46 & 0.73 & 0.66 & 1.23 & 0.86 \\
2251+134 & 0.595 & 0.06 & 0.16 & 0.80 & 0.96 & 0.85\\
& & & \\ \hline 
\end{tabular}
\end{center}
{\bf Columns Explanation:} {1=Source IAU name; 2=5 GHz flux
in Jy (Cheung et al. 2003); 3=Dereddened 5500\AA\ flux in mJy; 4=1~keV
{\it Chandra} flux in $\mu$Jy from best-fit in Table 4 or 5;
5-7=Spectral Indices for radio-optical, optical-X-ray, and radio-X-ray
($F_{\nu}\propto\nu^{-\alpha}$).}
\end{table}

\subsection{The low-power radio galaxy 0836+299.} 

The nearby FRI radio galaxy 0836+299 has interesting properties which
distinguish it from the other quasars in our sample. This is the only
source with evidence for diffuse X-ray emission on the galaxy's halo
scale, with X-ray luminosities consistent with other FRIs (Worrall et
al. 2001).

Interestingly, we find evidence for excess absorption from the core,
with column density N$_H \sim 5 \times 10^{23}$ cm$^{-2}$\ (Table
5). This is at odds with the recent claim (Hardcastle et al. 2002 and
references therein) that FRIs lack intrinsic X-ray
absorption. Suggested scenarios explain the lack of column density by
considering that FRIs do not have a molecular torus obscuring the
isotropic core emission, or that the inner jet contributes to the
X-ray emission on scales larger than the torus. Corroborating evidence
for these scenarios was provided by optical studies with {\it HST} of
a sample of FRIs (Chiaberge et al. 2000), showing a high detection
rate of nuclear point sources.

Our findings for 0836+299 (and for other FRIs studied at X-rays by our
group) show that at least {\it some} FRIs possess excess X-ray
absorption. The optical extinction implied by the X-ray column density
is A$_V \sim 240$ mag, using Galactic gas-to-dust ratios (A$_V \sim 5
\times 10^{-22}$ N$_H$ mag/cm$^{-2}$).  A nuclear dust lane has been observed
in 0836+299 in ground-based optical images (van Breugel et al. 1986)
and in our {\it HST} images (Scarpa et al. 2003). In fact, the source
has a flat $\alpha_{ox}$ index and is a luminous IR source detected
with IRAS at 25 and 60$\mu$m (Golombek et al. 1988), indicating the
presence of dust.

The amount of reddening for the optical nuclear light is A$_V=1.2$ mag
(van Breugel et al. 1986). This is much less than the reddening
derived from our X-ray column. Therefore, it seems likely that at
least part of the X-ray absorber originates in a different component
than the gas responsible for the optical reddening (unless that
gas-to-dust ratio is much different from Galactic; e.g., Maiolino et
al. 2001).  The origin of the X-ray absorbing medium in 0836+299, as
in other FRIs studied by us (Sambruna et al. 2003), is not clear. For
large inclinations ($\; \buildrel > \over \sim \;$80\degr), a natural
possibility is the molecular torus. Indeed, the column density we
measure for 0836+299 is consistent with edge-on Seyfert 2s (Risaliti
et al. 1999), and the small value of $R_i$ in Table 1 for this galaxy
suggests large inclinations. Alternatively, it was proposed that the
X-ray absorbing medium is photoionized gas inside the torus
(Weingartner \& Murray 2002).

Independent of the origin of the X-ray absorber in 0836+299, the
existence of intrinsic obscuration of FRIs remains a controversial
topic, at best. A larger sample of FRIs observed with {\it Chandra}
and {\it XMM-Newton} is needed to address this issue, which would
complement ongoing {\it HST} studies by other groups (Chiaberge et
al. 2002).\vspace{0.25cm}

\noindent\appendix{\bf Appendix A: Previous X-ray observations}

\noindent In Table A, we collected the results from previous X-ray
observations of the sources of our sample, for comparison with the
{\it Chandra} spectral results in Table 4-5.

\begin{table*}
\begin{center}
\begin{tabular}{llllllll}
\multicolumn{8}{l}{{\bf Table A: Earlier Spectral Fits for X-ray Continua}} \\
\multicolumn{8}{l}{   } \\ \hline \hline
& & & \\
\multicolumn{1}{c}{Source} &\multicolumn{1}{c}{N$_{H}$} & \multicolumn{1}{c}{Photon Index} & \multicolumn{1}{c}{Other Parameters} & \multicolumn{1}{c}{$\chi^{2}_r$/dof} & \multicolumn{1}{c}{$F_{1\ keV}$} & \multicolumn{1}{c}{Instrument/Band} & \multicolumn{1}{c}{References} \\
\multicolumn{1}{c}{(1)} &\multicolumn{1}{c}{(2)} & \multicolumn{1}{c}{(3)} & \multicolumn{1}{c}{(4)} & \multicolumn{1}{c}{(5)} & \multicolumn{1}{c}{(6)} & \multicolumn{1}{c}{(7)} & \multicolumn{1}{c}{(8)}\\ 
& & & \\ \hline
& & & \\
0405$-$123
& 7.20 & $\Gamma = 1.76^{+0.09}_{-0.10}$ & $\cdots$ & 1.00/160 & 0.90 & {\it ASCA} (2--10) & D01 \\

& 3.87 & $\Gamma = 2.36^{+0.05}_{-0.06}$ & $\cdots$ & 0.91/19 & 1.01$^a$ & {\it ROSAT} (0.1--2.4) & S98 \\

& 3.7 & $\Gamma = 2.33 \pm 0.06$ & $\cdots$ & 1.12/79 & 1.16$\pm 0.05$ & {\it ROSAT} (0.1--2.4)& S97 \\

0836+299 
& 4.1 & $\Gamma = 1.7$ & $\cdots$ & $\cdots$ & 0.03$^c$ & {\it Einstein} (0.2--4.5) & BM91 \\

0838+133 
& 16.5 &  $\Gamma = 1.22^{+0.06}_{-0.05}$ & $\cdots$ & 1.11/182 & 0.11$\pm$ 0.01 & {\it Chandra} (0.5--8) & B01 \\

1055+018 
& 3.45 & $\Gamma = 2.20 \pm 0.16$ & $\cdots$ & 0.99/14 & 0.21$\pm 0.03$ & {\it ROSAT} (0.1--2.4) & S97 \\

& 1.62 & $\Gamma =
1.64^{+0.47}_{-0.43}$ & $\cdots$ & 0.85/10 & 0.28 & {\it
ROSAT} (0.1--2.4) & S98 \\

1150+497 
& 2.1 & $\Gamma = 2.14 \pm 0.05$ & $\cdots$ & 0.93/67 & 0.55$\pm 0.03$ & {\it ROSAT} (0.1--2.4) & S97 \\

1354+195 
& 5.0 & $\Gamma = 1.5$ & $\cdots$ & $\cdots$ & 1.19 $\pm 0.36^c$ & {\it Einstein} (0.2--4.5) & B87 \\

1510$-$089 & 8.7 & $\Gamma =
1.60^{+0.39}_{-0.35}$ & $\cdots$ & 0.44/12 & 0.87$\pm 0.23$ & {\it
  EXOSAT} (2--10) & G95 \\

& 7.6 & $\Gamma = 1.89^{+0.16}_{-0.17}$ &
$\cdots$& 1.30/26 & 0.66$\pm 0.05$ & {\it ROSAT} (0.1--2.4) & S97 \\

& 6.80 & $\Gamma = 1.92 \pm 0.15$ & $\cdots$ & 0.97/28 & 0.72 & {\it ROSAT} (0.1--2.4) & S98 \\

& 9.8 & $\Gamma = 1.4\pm 0.1$ & $E_{\ell} = 4.95\pm 0.25$ & 0.97/50 & 0.90 & {\it EXOSAT} (2--10) & S90 \\
& & & $\sigma$=0.1\\

& 7.8 & $\Gamma_{SOFT} = 2.65 \pm 0.63$ & E$_0=1.3\pm0.3$& 0.68/63 & 0.53$^b$ & {\it BeppoSAX} (0.4--100) & T00 \\ 
& & $\Gamma_{HARD} = 2.65 \pm 0.63$ \\

1641+399 
& 0.9 & $\Gamma = 1.98 \pm
0.08$ & $\cdots$ & 1.50/44 & 0.38$\pm 0.03$ & {\it ROSAT} (0.1--2.4) & S97 \\

& 1.0 & $\Gamma = 1.7$ & $\cdots$& $\cdots$ &
0.47$^c$ & {\it Einstein} (0.2--4.5) & BM91\\

1928+738 
& 11.67 & $\Gamma = 2.33^{+0.21}_{-0.17}$ & $\cdots$ & 0.83/82 & 1.17$\pm 0.08$ & {\it ROSAT} (0.1--2.4) & S97 \\

& 10.4 & $\Gamma = 2.1\pm 0.3$ & $\cdots$ & 1.05/30 & 1.54$^a$ & {\it EXOSAT} (2--10) & L92 \\

& 13.0 & $\Gamma = 2.25\pm 0.48$ & $\cdots$ & 0.97/29 & 0.52$\pm 0.08$ & {\it EXOSAT} (2--10) & G95  \\

& & & \\\hline 
\end{tabular}
\end{center}
{\bf Columns Explanation:} {1=Source IAU name; 2-4=Model
and Parameter Values for fits from literature: N$_H$=absorption column
density, ($\times 10^{20}$ cm$^{-2}$); $E_{\ell}$=gaussian line
energy, in keV; $\sigma$=gaussian line width, in keV; E$_0$=Break
Energy between $\Gamma_{SOFT}$ and $\Gamma_{HARD}$ in keV; 5=Reduced
$\chi^2$ and degrees of freedom (dof); 6=Flux density at 1~keV in
$\mu$Jy; 7=X-ray Instrument used for measurement / Sensitivity for the
instrument in keV; 8=Reference for the reported spectra. B87=Biermann
et al. (1987); BM91=Bloom \& Marscher (1991); B01=Brunetti et
al. (2001); D01=Donato et al. (2001); G95=Ghosh et al. (1995);
L92=Lawson et al. (1992); S97=Sambruna (1997); S98=Siebert et
al. (1998); S90=Singh et al. (1990); T00=Tavecchio et al. (2000).}

\noindent{\bf Notes:} {$a$=Monochromatic flux was calculated by us;
$b$=Monochromatic flux density given at 2~keV; $c$=values for N$_{H}$
and $\Gamma$ are fixed.}
\end{table*}

\begin{acknowledgements}

We are grateful to M. Gliozzi and D. Donato for their assistance with
the spatial analysis. JKG and RMS are funded by NASA grants
NAS8-39073 and GO 1-2110A, the latter operated by AURA, Inc., under
NASA contract NAS 5-26555. CMU acknowledges support from LTSA grant
NAG5-9327.  CCC acknowledges that radio astronomy at Brandeis
University is supported by the NSF.  This research has made use of the
NASA/IPAC Extragalactic Database (NED) which is operated by the Jet
Propulsion Laboratory, California Institute of Technology, under
contract with the National Aeronautics and Space Administration.

\end{acknowledgements}

\end{document}